\documentclass[conference]{IEEEtran}
\IEEEoverridecommandlockouts
\usepackage{cite}
\usepackage{amsmath,amssymb,amsfonts}
\usepackage{algorithmic}
\usepackage{algorithm}
\usepackage{graphicx}
\usepackage{textcomp}
\usepackage{xcolor}
\usepackage{booktabs}
\DeclareRobustCommand{\rev}[1]{#1}
\def\BibTeX{{\rm B\kern-.05em{\sc i\kern-.025em b}\kern-.08em
    T\kern-.1667em\lower.7ex\hbox{E}\kern-.125emX}}
\begin{document}
\bstctlcite{IEEEexample:BSTcontrol}

\title{Robust Optimal Network Reduction for Distribution Systems
\thanks{\rev{This work was supported in part by a grant from the Leahy Institute for Rural Partnerships at the University of Vermont.}}
}

\author{\IEEEauthorblockN{Omid Mokhtari\IEEEauthorrefmark{1}, James Bagrow\IEEEauthorrefmark{2}, Mads R. Almassalkhi\IEEEauthorrefmark{1}, and  Samuel Chevalier\IEEEauthorrefmark{1}}
\IEEEauthorblockA{\IEEEauthorrefmark{1}\textit{Department of Electrical and Biomedical Engineering}, \textit{University of Vermont}, Burlington, VT 05405, USA\\
\IEEEauthorrefmark{2}\textit{Department of Mathematics and Statistics}, \textit{University of Vermont}, Burlington, VT 05405, USA\\
\{omid.mokhtari, james.bagrow, malmassa, schevali\}@uvm.edu}
}

\maketitle

\begin{abstract}
Network reduction replaces large networks with smaller equivalents to speed up power flow and optimization studies. Optimal Kron-based Reduction of Networks (Opti-KRON) finds the optimal set of nodes to faithfully represent a large-scale radial network for representative loading scenarios. Under unseen nodal injections, however, its accuracy can drop.
This paper extends Opti-KRON to uncertain nodal injections using robust uncertainty sets and a Wasserstein ambiguity set, with voltage magnitude error guarantees. 
To improve scalability, we develop a presolve screening step to eliminate redundant constraints and variables while preserving the feasible set. 
We also adopt adversarial constraint generation to reduce memory requirements by iteratively adding worst-case nodal injections.
On a real 8,387-node feeder at a threshold of 0.001~p.u., the reduction removes 49.9\% of the nodes. For one uncertainty set, adversarial constraint generation cuts the required memory by a factor of more than 18.
\end{abstract}


\section{Introduction}
The computational burden of power flow and optimization problems in distribution systems increases with network size. Network reduction lowers this burden by replacing a large feeder with a smaller equivalent network. However, reducing network size introduces modeling errors.

Various network reduction methods have been developed to reduce distribution networks while preserving the accuracy of nodal voltages~\cite{Dorfler2013,Pecenak2019}.
Optimal Kron-based Reduction of Networks (Opti-KRON) formulates network reduction as a mixed-integer linear program (MILP) that selects the nodes to keep. It minimizes the number of kept nodes while keeping the voltage magnitude errors within a threshold for a set of representative loading scenarios \cite{Chevalier2022,Mokhtari2026,MokhtariPSCC2026}. Kron reduction eliminates the other nodes from the nodal admittance equations and gives an equivalent network on the kept nodes~\cite{Dorfler2013}.

The accuracy of Opti-KRON depends on the loading scenarios considered during the reduction process. As nodal injections vary due to changes in demand and generation, operating conditions outside these scenarios may result in voltage magnitude errors above the threshold. Robust optimization accounts for such variations by considering a set of possible realizations of uncertain nodal injections~\cite{Roald2023}. However, the resulting robust optimization problems can be computationally prohibitive for large feeders.
To address these challenges, this paper makes the following contributions:
\begin{itemize}
\item \textbf{Robust optimal network reduction:} We extend Opti-KRON to uncertain nodal injections through robust uncertainty sets and a Wasserstein ambiguity set.
Under LinDistFlow, the voltage magnitude error is linear in the injections for a given reduction, which allows exact reformulations of the error constraints.
\item \textbf{Scalable solution methods:}
We develop screening rules based on voltage differences within candidate clusters and bounds on error coefficients to eliminate infeasible node assignments and redundant error constraints.
We also adopt adversarial constraint generation to avoid constructing large robust counterparts by iteratively adding worst-case nodal injections.

\item \textbf{Validation on realistic networks:} We validate and compare methods on large-scale distribution feeders. 
\end{itemize}

\section{Robust Opti-KRON}

\subsection{Kron-Based Reduction and Voltage Magnitude Error}
\begin{figure}[t]
\centering
\includegraphics[width=\columnwidth]{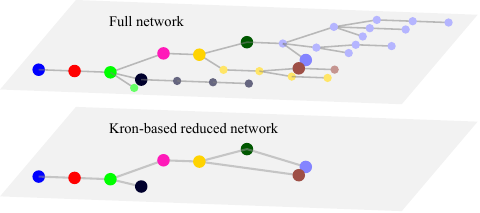}
\caption{A 30-node radial feeder (top) and its reduced network with nine super-nodes (bottom), whose lines follow from Kron reduction. Each color represents a cluster of nodes.}
\label{fig:example}
\end{figure}

Opti-KRON reduces a network by keeping a subset of its nodes, called super-nodes~\cite{Mokhtari2026}. Each of the other nodes, called reduced nodes, is assigned to one super-node. A super-node and its reduced nodes form a cluster, and the injections of the cluster are aggregated at the super-node. The reduced nodes then carry no injection.
The voltage of a super-node is used as an estimate for the voltages of its reduced nodes. The voltage magnitude error of a node is the difference between this estimate and the node's voltage magnitude in the full network. Fig.~\ref{fig:example} shows the clusters of a 30-node feeder and the resulting reduced network.

For a radial feeder with node set $\mathcal{V}=\{1,\dots,n\}$ and slack node $s$, the assignment is a binary matrix $A\in\{0,1\}^{n\times n}$, where $A_{ij}=1$ if node $j$ is assigned to super-node $i$, and $A_{ii}=1$ marks a super-node. The candidate set $\mathcal{C}$ contains the pairs $(i,j)$ of nodes that are at most $H$ edges apart, where $H$ is the hop limit. The hop limit keeps the clusters small and removes distant candidate pairs from the problem. Node $j$ can be assigned to super-node $i$ only if $(i,j)\in\mathcal{C}$. A valid assignment satisfies
\begin{subequations}\label{eq:assign}
\begin{align}
&\textstyle\sum_{i} A_{ij}=1,\ \ A_{ij}\le A_{ii},\ \ A_{ss}=1, \quad \forall i,j, \label{eq:assign_a}\\
&A_{ij}\le A_{ik}, \quad \forall (i,j)\in\mathcal{C},\ k\in\mathcal{P}_{ij}, \label{eq:assign_b}\\
&A_{ij}=0, \quad \forall (i,j)\notin\mathcal{C}. \label{eq:assign_c}
\end{align}
\end{subequations}
Constraints~\eqref{eq:assign_a} assign every node to one super-node and keep the slack node. In~\eqref{eq:assign_b}, $\mathcal{P}_{ij}$ is the set of nodes on the path between $i$ and $j$, so every cluster is a connected subtree.

We model the voltages with the LinDistFlow equations~\cite{Baran1989}. Let the network be rooted at the slack node $s$, with edge set $\mathcal{E}$ of $n-1$ edges and vectors of edge resistances $r$ and reactances $x$. Let $p,q\in\mathbb{R}^n$ be the active and reactive nodal injections, with generation positive. For voltage magnitudes $v\in\mathbb{R}^n$ near 1~p.u., we use the approximation
\begin{align}
v &= v_0\mathbf{1} + K_P\,p + K_Q\,q, \label{eq:lindistflow}\\
K_P &= \Gamma^\top \mathrm{diag}(r)\,\Gamma,\quad K_Q = \Gamma^\top \mathrm{diag}(x)\,\Gamma, \label{eq:sens}
\end{align}
where $v_0$ is the slack voltage magnitude and $\Gamma\in\{0,1\}^{(n-1)\times n}$ has $\Gamma_{ej}=1$ if node $j$ is downstream of edge $e$. Thus $K_{P,jk}$ is the resistance shared by the paths from $s$ to $j$ and to $k$. \rev{For three-phase networks, we use the phase-coupled LinDistFlow model of~\cite{Gan2014} converted to voltage magnitudes as in~\eqref{eq:lindistflow}, with one row per existing node-phase pair and with $K_P$ and $K_Q$ built from the phase impedance matrices of the lines. A node can be assigned only to a super-node that has all its phases. Injections are aggregated phase by phase, and the voltage magnitude error is defined for each phase.}

Under the aggregated injections $Ap$ and $Aq$, the voltage magnitudes become $\tilde{v} = v_0\mathbf{1} + K_P A p + K_Q A q$; the entries of $\tilde{v}$ at the super-nodes are the voltage magnitudes of the reduced network. The voltage magnitude error of node $j$ assigned to super-node $i$ is then
\begin{align}
e_{ij}(A,u) &= \tilde{v}_i - v_j = c_{ij}(A)^\top u, \label{eq:error}\\
c_{ij}(A) &= \begin{bmatrix} (K_PA)_{i,:}^\top-(K_P)_{j,:}^\top \\ (K_QA)_{i,:}^\top-(K_Q)_{j,:}^\top \end{bmatrix}, \label{eq:coef}
\end{align}
where $u=[p;q]$. For a fixed assignment, the error is linear in $u$, and its coefficients are affine in $A$. In~\cite{Mokhtari2026}, the error was computed from voltages and currents of representative AC power flow solutions. Here, \eqref{eq:error} is a LinDistFlow approximation of the voltage magnitude error. The robust formulations below are exact for this approximate error, not for the voltage magnitude error of the AC power flow.

\subsection{Robust Reduction Problem}
Let $\bar{E}$ be the voltage magnitude error threshold. The robust reduction problem finds the assignment with the fewest super-nodes such that every error stays within $\bar{E}$ for every injection in an uncertainty set $\mathcal{U}$:
\begin{subequations}\label{eq:robust}
\begin{align}
&\min_{A}  \  \textstyle\sum_{i} A_{ii} \label{eq:robust_obj}\\
\text{s.t.} \ & \max_{u\in\mathcal{U}}\, c_{ij}(A)^\top u \le \bar{E} + M_{ij}(1-A_{ij}),  \forall (i,j)\in\mathcal{C}, \label{eq:robust_max}\\
& \min_{u\in\mathcal{U}}\, c_{ij}(A)^\top u \ge -\bar{E} - M_{ij}(1-A_{ij}),  \forall (i,j)\in\mathcal{C}, \label{eq:robust_min}\\
& \eqref{eq:assign}\text{--}\eqref{eq:coef}. \label{eq:robust_struct}
\end{align}
\end{subequations}
When node $j$ is assigned to super-node $i$ ($A_{ij}=1$), \eqref{eq:robust_max} and~\eqref{eq:robust_min} keep its voltage magnitude error between $-\bar{E}$ and $\bar{E}$ for every injection in $\mathcal{U}$. When $A_{ij}=0$, the constant $M_{ij}\ge 0$ makes both constraints inactive for every assignment that satisfies~\eqref{eq:assign}. If $\mathcal{U}$ is a finite set of loading scenarios, \eqref{eq:robust_max} and~\eqref{eq:robust_min} enforce the LinDistFlow error bound at every scenario, as in scenario-based Opti-KRON.

\subsection{Uncertainty Representation}\label{sec:sets}
We build the uncertainty sets from $N$ training samples $u^1,\dots,u^N$ of the nodal injections. A coordinate of $u$ is active if it is nonzero in at least one sample, and $\mathcal{S}$ is the set of active coordinates. All other coordinates are zero in every set. Table~\ref{tab:sets} lists the uncertainty sets and the worst case of a linear function $c^\top u$ over each set. For the robust uncertainty sets, replacing the maximum in~\eqref{eq:robust_max} by this worst case gives an exact reformulation, called the robust counterpart~\cite{BenTal2009}. \rev{For~\eqref{eq:robust_min}, the robust counterpart uses Table~\ref{tab:sets} with $-c_{ij}(A)$.}

\begin{table}[t]
\centering
\caption{Uncertainty sets and the worst case of $c^\top u$}
\label{tab:sets}
\setlength{\tabcolsep}{3.5pt}
\begin{tabular}{lll}
\toprule
Set & Definition & Worst case of $c^\top u$\\
\midrule
Box & $\ell\le u\le h$ & $c^\top\bar{u}+\delta^\top|c|$\\
Convex hull & $\mathrm{conv}\{u^1,\dots,u^N\}$ & $\max_{k} c^\top u^k$\\
Ellipsoid & $u_{\mathcal{S}}=\mu+Ly,\ \|y\|_2\le\Omega$ & $c_{\mathcal{S}}^\top\mu+\Omega\|L^\top c_{\mathcal{S}}\|_2$\\
Wasserstein & $W(\mathbb{Q},\hat{\mathbb{P}}_N)\le\rho$ & $\sup_{\mathbb{Q}\in\mathcal{Q}}\mathrm{CVaR}^{\mathbb{Q}}_{1-\epsilon}(c^\top u)$\\
\bottomrule
\end{tabular}
\end{table}

The box bounds each active coordinate between its smallest and largest sampled values, $\ell$ and $h$. Its midpoint and half-width are $\bar{u}=(\ell+h)/2$ and $\delta=(h-\ell)/2$.
\rev{In the worst case, each coordinate of $u$ is at its upper bound where the matching coordinate of $c$ is positive, and at its lower bound otherwise.} 

The convex hull contains all convex combinations of the samples, and it lies inside the box. Since the error is linear in $u$, its maximum over the convex hull is reached at one of the samples. The robust counterpart therefore enforces the bound at every training sample.

The ellipsoid describes how the injections vary together. Its center $\mu$ is the mean of the training samples on $\mathcal{S}$, and $L$ is the Cholesky factor of the covariance $\Sigma=LL^\top\succ 0$ of the active coordinates. In this paper, we consider a radius $\Omega$ equal to a quantile of the scaled distances $\|L^{-1}(u_{\mathcal{S}}-\mu)\|_2$ of the training samples~\cite[Ch.~2]{BenTal2009}.
Because of the Euclidean norm term in Table~\ref{tab:sets}, the robust counterpart is a second-order cone constraint, and problem~\eqref{eq:robust} is a mixed-integer second-order cone program (MISOCP).

The Wasserstein ambiguity set $\mathcal{Q}$ contains every distribution $\mathbb{Q}$ of the active coordinates with $W(\mathbb{Q},\hat{\mathbb{P}}_N)\le\rho$, where $W$ is the type-1 Wasserstein distance with a norm $\|\cdot\|$ as transport cost, and $\hat{\mathbb{P}}_N$ is the empirical distribution of the training samples~\cite{EsfahaniKuhn2018}. For this set, the worst case is the largest conditional value-at-risk (CVaR) of $c^\top u$ at level $1-\epsilon$, with $\epsilon\in(0,1]$. If the active coordinates can vary freely, $\sup_{\mathbb{Q}\in\mathcal{Q}}\mathrm{CVaR}^{\mathbb{Q}}_{1-\epsilon}(c^\top u)=\max_{u\in\mathcal{U}_W}c^\top u$~\cite{EsfahaniKuhn2018}, where
{\small
\begin{equation}
\mathcal{U}_W=\{\textstyle\sum_k\theta_k u^k+d:\ 0\le\theta_k\le\frac{1}{N\epsilon}, \|d\|\le\frac{\rho}{\epsilon},  \sum_k \theta_k = 1 \}. \label{eq:uw}
\end{equation}}
In this paper, we consider the worst-case CVaR of each error separately for each node and sign~\cite{Roald2023}. As for $\Omega$, we consider a radius $\rho/\epsilon$ equal to a quantile of the distances of the training samples to the subset of $\mathcal{U}_W$ with $d=0$~\cite{EsfahaniKuhn2018}.

\section{Scalable Solution Methods}
Constructing and solving the robust counterpart of~\eqref{eq:robust} can be computationally expensive for large feeders and data libraries. We address this challenge in two ways: a presolve screening step to reduce the problem size and adversarial constraint generation to solve~\eqref{eq:robust} through a sequence of smaller scenario-based MILPs.

\subsection{Constraint Screening}\label{sec:prep}
A candidate pair $(i,j)$ is infeasible if no feasible solution of~\eqref{eq:robust} has $A_{ij}=1$. An error constraint of pair $(i,j)$ is redundant if it holds for every valid assignment with $A_{ij}=1$. Infeasible pairs and redundant error constraints can be removed without changing the feasible set or the optimal value of~\eqref{eq:robust}. We develop two screening tests based on voltage magnitude differences and bounds on the error coefficients.

The first test exploits voltage magnitude differences between nodes that must share a super-node. If $A_{ij}=1$, constraint~\eqref{eq:assign_b} assigns the nodes in $\mathcal{P}_{ij}$ to super-node $i$, so all nodes in $\mathcal{F}_{ij}=\mathcal{P}_{ij}\cup\{i,j\}$ share super-node $i$. For any two nodes $a,b\in\mathcal{F}_{ij}$, the super-node voltage cancels in the difference of their errors:
\begin{equation}
e_{ia}(A,u)-e_{ib}(A,u)=v_b-v_a=d_{ab}^{\top}u, \label{eq:spread}
\end{equation}
where $d_{ab}$ follows from~\eqref{eq:lindistflow} and is independent of $A$. If both errors are within $\pm\bar{E}$, their difference cannot exceed $2\bar{E}$ in magnitude. Hence, pair $(i,j)$ is infeasible if
\begin{equation}
\max_{u\in\mathcal{U}}|d_{ab}^{\top}u|>2\bar{E}
\end{equation}
for some $a,b\in\mathcal{F}_{ij}$ \rev{on the same phase}. This maximum follows from Table~\ref{tab:sets}, applied to $d_{ab}$ and $-d_{ab}$.

The second test exploits the structure of the error coefficients in~\eqref{eq:coef}. Although $c_{ij}(A)$ depends on the whole assignment, each entry depends only on the super-node of a single node. In particular, the entry multiplying $p_k$ is
\begin{equation}
c_{ij,k}(A)=(K_PA)_{ik}-K_{P,jk}=K_{P,ib}-K_{P,jk},
\end{equation}
where $b$ is the super-node of node $k$ ($A_{bk}=1$). Let $\mathcal{C}_k=\{b:(b,k)\in\mathcal{C}\}$ denote the remaining candidate super-nodes of node $k$. Since $b\in\mathcal{C}_k$, this entry satisfies
\begin{equation}
c_{ij,k}(A)\in\left[\min_{b\in\mathcal{C}_k}K_{P,ib}-K_{P,jk},\max_{b\in\mathcal{C}_k}K_{P,ib}-K_{P,jk}\right]. \label{eq:interval}
\end{equation}
Analogous bounds hold for the entries multiplying $q_k$, with $K_Q$ replacing $K_P$. Moreover, if $A_{ij}=1$, every node $k\in\mathcal{F}_{ij}$ is assigned to super-node $i$, so the entries multiplying $p_k$ and $q_k$ are known exactly. We use these intervals to compute a lower bound $\underline{h}_{ij}$ and an upper bound $\overline{h}_{ij}$ of the worst-case error $\max_{u\in\mathcal{U}} c_{ij}(A)^\top u$. Both bounds hold for every assignment with $A_{ij}=1$, and they lead to two conclusions:
\begin{subequations}\label{eq:screen}
\begin{align}
&\underline{h}_{ij}>\bar{E}\ \Rightarrow\ \text{pair $(i,j)$ is infeasible},\\
&\overline{h}_{ij}\le\bar{E}\ \Rightarrow\ \text{constraint~\eqref{eq:robust_max} is redundant}.
\end{align}
\end{subequations}
The same test with $-c_{ij}(A)$ applies to~\eqref{eq:robust_min}.

For the box, each term $c_m\bar{u}_m+|c_m|\delta_m$ of the worst case in Table~\ref{tab:sets} depends on one entry. The sum of the smallest values of these terms over their intervals is the lower bound, and the sum of the largest values is the upper bound. For the convex hull, the bounds follow from bounds of $c_{ij}(A)^\top u^k$ at each sample, and each sample constraint is tested on its own. \rev{For the Wasserstein ambiguity set, we maximize the weighted sum of the sample bounds over the weights in~\eqref{eq:uw} and bound the additional term $(\rho/\epsilon)\|c_{ij,\mathcal{S}}(A)\|_*$ over the coefficient intervals, where $\|\cdot\|_*$ is the dual norm of $\|\cdot\|$. For the ellipsoid, we bound the linear term and each entry of $L^\top c_{ij,\mathcal{S}}(A)$ to obtain only an upper bound on the worst-case error in Table~\ref{tab:sets}.}

Without the condition $A_{ij}=1$, the intervals~\eqref{eq:interval} and the upper bounds hold for every assignment. We choose $M_{ij}$ at least as large as each of these upper bounds, for both signs, minus $\bar{E}$. Each infeasible pair removes a candidate super-node, which can narrow the intervals of other pairs. In addition, by~\eqref{eq:assign_b}, if pair $(i,k)$ is infeasible and $k\in\mathcal{P}_{ij}$, pair $(i,j)$ is infeasible too. We therefore repeat the second test and this path rule for a few passes, until no new infeasible pair is found or the pass limit is reached.

\subsection{Adversarial Constraint Generation}
After screening, problem~\eqref{eq:robust} can also be solved by adversarial constraint generation, a cutting-set method~\cite{MutapcicBoyd2009} in which an adversary searches $\mathcal{U}$ for the worst-case injections of a given reduction, as in robust optimization~\cite{Roald2023}. The method enforces the error bounds only at a finite set of these injections. It alternates between the adversary and a master problem, which is~\eqref{eq:robust} with $\mathcal{U}$ replaced by this finite set $\mathcal{U}^t\subset\mathcal{U}$. The error constraints of the master problem are linear, as for the convex hull. Adversarial constraint generation therefore avoids the robust counterpart, and it can help when this counterpart grows with the number of samples or is conic~\cite{BertsimasDunningLubin2016}.

Since $\mathcal{U}^t\subset\mathcal{U}$, the optimal value of the master problem is a lower bound on that of~\eqref{eq:robust}, and its reduction $A^t$ may violate the bounds at injections outside $\mathcal{U}^t$. In $A^t$, each node $j$ is assigned to one super-node $i$, and we write $e^t_j(u)=e_{ij}(A^t,u)$ for its error. For each node, the adversary finds the injection that causes the largest violation for $A^t$. The largest violation over all nodes is
\begin{equation}
\Delta^t=\max_{j}\ \max_{u\in\mathcal{U}}\ |e^t_j(u)|-\bar{E}. \label{eq:pernode}
\end{equation}
For a fixed node, the error is $c^\top u$ with $c=c_{ij}(A^t)$. Its largest magnitude over $\mathcal{U}$ is the larger of the maxima of $c^\top u$ and $-c^\top u$, given by Table~\ref{tab:sets}. The set $\hat{\mathcal{U}}^t$ contains the worst-case injections of at most $K$ violated nodes, those with the largest violations. 
The added injections violate the error bound for $A^t$, so the master problem cannot return $A^t$ again. If $\Delta^t\le 0$, $A^t$ is feasible for~\eqref{eq:robust}, and by the lower bound it is optimal. Since the number of assignments is finite, Algorithm~\ref{alg:acg} stops after finitely many iterations.


\begin{algorithm}[t]
\caption{Adversarial constraint generation}
\label{alg:acg}
\begin{algorithmic}[1]
\REQUIRE $\mathcal{U}$, $\bar{E}$, $\mathcal{U}^0\subset\mathcal{U}$, $K$
\FOR{$t=0,1,2,\dots$}
\STATE $A^t\gets$ optimal solution of~\eqref{eq:robust} with $\mathcal{U}^t$ in place of $\mathcal{U}$
\STATE $(\Delta^t,\hat{\mathcal{U}}^t)\gets$ adversary~\eqref{eq:pernode} for $A^t$
\IF{$\Delta^t\le 0$}
\RETURN $A^t$
\ENDIF
\STATE $\mathcal{U}^{t+1}\gets\mathcal{U}^t\cup\hat{\mathcal{U}}^t$
\ENDFOR
\end{algorithmic}
\end{algorithm}

\section{Case Studies}
We test the methods on South Alburg, a real three-phase feeder in Vermont with 8,387 nodes, and on the IEEE European LV test feeder with 907 nodes~\cite{Schneider2018}. South Alburg has one month of historical hourly load data: 29 days are training samples, and one day is the test day. The European LV feeder has one day of one-minute profiles, and all 1,440 samples are training samples. We use $H=5$, $\bar{E}=0.001$~p.u., $\epsilon=0.01$, and the $\ell_1$, $\ell_2$ and $\ell_\infty$ norms as transport cost. \rev{On both feeders, we also vary $\bar{E}$ from 0.001 to 0.005~p.u.} The radii $\Omega$ and $\rho/\epsilon$ are the 95\% quantiles of the distances defined in Section~\ref{sec:sets}. The covariance of the ellipsoid is estimated by principal orthogonal complement thresholding with at most five factors~\cite{Fan2013}. Each set is solved with the robust counterpart of~\eqref{eq:robust}, called the direct model, and with Algorithm~\ref{alg:acg} with $K=3$. \rev{As a baseline, we use scenario-based Opti-KRON~\cite{Mokhtari2026}, solving~\eqref{eq:robust} with $\mathcal{U}$ equal to the two training samples of largest and smallest total net active injection. We initialize $\mathcal{U}^0$ with these samples if they lie in $\mathcal{U}$.} We use Gurobi with a zero optimality gap and a 24-h time limit. \rev{Computation times include constraint screening and solving. All runs use eight threads at the Vermont Advanced Computing Center.} 

\subsection{Reduction for Each Set}
\begin{figure}[t]
\centering
\includegraphics[width=\columnwidth]{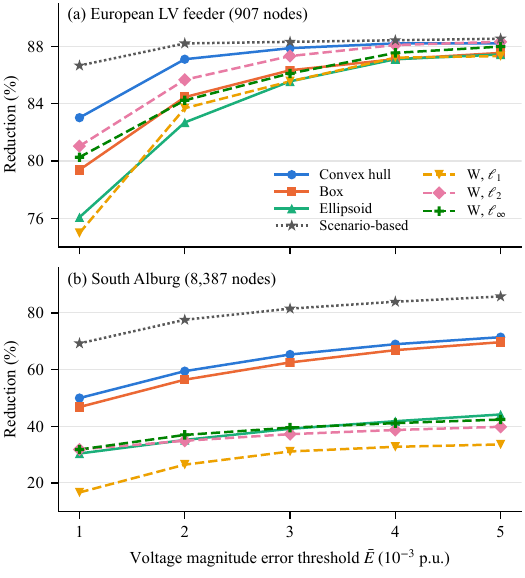}
\caption{\rev{Reduction versus the voltage magnitude error threshold for (a) the European LV feeder and (b) South Alburg. W, $\ell_p$: Wasserstein ambiguity set with the $\ell_p$ norm as transport cost.}}
\label{fig:tradeoff}
\end{figure}
\rev{Fig.~\ref{fig:tradeoff}(a) shows the reduction of the European LV feeder. At $\bar{E}=0.001$~p.u., it ranges from 75.0\% for the $\ell_1$ Wasserstein set to 83.0\% for the convex hull. The convex hull lies inside the box, so the box cannot give a larger reduction. The Wasserstein sets are not nested with the box, and the $\ell_2$ and $\ell_\infty$ sets reduce the feeder more than the box. As $\bar{E}$ grows, all curves, including the baseline, approach 88.6\%, the largest reduction that the structural constraints alone allow.}

\rev{On South Alburg, the reduction also increases with $\bar{E}$, but, unlike on the European LV feeder, the gap between the sets does not close (Fig.~\ref{fig:tradeoff}(b)). At $\bar{E}=0.001$~p.u., the convex hull and the box reduce the feeder by 49.9\% and 46.8\%, against 16.6\% to 31.8\% for the ellipsoid and the Wasserstein sets. The box and the convex hull stay within the sampled range of each coordinate. In contrast, the ellipsoid and the Wasserstein sets use a single radius calibrated across all 5,614 active coordinates, while each voltage magnitude error is sensitive mainly to a subset of injections. The worst case of each error can exploit the entire radius in these directions. This allows larger deviations than those observed in the training samples.}

\rev{The scenario-based baseline reduces South Alburg by 69.3\% at the same threshold, but its voltage magnitude errors reach $102.7\bar{E}$ on the other training samples. The gap between the scenario-based and robust reductions illustrates the price of robustness~\cite{BertsimasSim2004}. On the European LV feeder, the gap is smaller: the baseline gives 86.7\% at $\bar{E}=0.001$~p.u., and the gap almost closes as $\bar{E}$ grows.}

\subsection{Computational Performance}
\begin{table}[t]
\centering
\caption{Computation time (s) with and without constraint screening}
\label{tab:prep}
\footnotesize
\setlength{\tabcolsep}{8pt}
\begin{tabular}{lcccc}
\toprule
 & \multicolumn{2}{c}{Direct} & \multicolumn{2}{c}{Algorithm~\ref{alg:acg}} \\
\cmidrule(lr){2-3}\cmidrule(lr){4-5}
Set & With & Without & With & Without \\
\midrule
Convex hull & 174 & OOM & 266 & 4,133 \\
Box & 4.1 & 64.4 & 9.9 & 2,473 \\
\bottomrule
\end{tabular}
\end{table}
On the European LV feeder, screening keeps 54\% to 70\% of the 11,227 candidate pairs at $\bar{E}=0.001$~p.u. and 89\% to 98\% at 0.005~p.u. Table~\ref{tab:prep} compares the computation time on this feeder with and without screening for $\bar{E}=0.001$~p.u. Where both runs finished, the run without screening takes about 16 to 250 times as long. The direct model of the convex hull without screening has 109 million constraints and runs out of memory (OOM) at 128~GiB.

On South Alburg, screening takes between 18~seconds (for the box) and 66~minutes (for the ellipsoid). It removes 54.4\% to 90.3\% of the binary variables as infeasible pairs and 91.7\% to 99.99\% of the error constraints. For the convex hull, the error constraints drop from 231 million to 2.9 million.

\begin{table}[t]
\centering
\caption{Computation time / peak memory (GiB) on South Alburg}
\label{tab:time}
\footnotesize
\setlength{\tabcolsep}{10pt}
\begin{tabular}{l c c}
\toprule
Set & Direct & Algorithm~\ref{alg:acg} \\
\midrule
Convex hull & 6.1~h / 95 & 8.8~h / 36 \\
Box & 4~min / 20 & 19~min / 21 \\
Ellipsoid & OOM / $>$480 & 8.4~h / 26 \\
Wasserstein, $\ell_1$ & 14~min / 20 & 16~min / 21 \\
Wasserstein, $\ell_2$ & 29~min / 26 & 29~min / 22 \\
Wasserstein, $\ell_\infty$ & 26~min / 46 & 32~min / 22 \\
\bottomrule
\end{tabular}
\end{table}
The direct model and Algorithm~\ref{alg:acg} give the same number of super-nodes where both runs finished, and the direct model is at least as fast (Table~\ref{tab:time}). For the box, screening removes most error constraints, and the robust counterpart does not grow with the number of samples. For the ellipsoid, the direct model exceeds the memory limit of 480~GiB while it is built, and only Algorithm~\ref{alg:acg} proves the minimal number of super-nodes. In the finished runs, Algorithm~\ref{alg:acg} needs at most 36~GiB, against 95 and 46~GiB for the direct convex-hull and $\ell_\infty$ Wasserstein models. \rev{As $\bar{E}$ grows, screening removes fewer candidate pairs, and the direct models need more memory: at $\bar{E}=0.002$~p.u., the direct convex-hull and $\ell_1$ and $\ell_2$ Wasserstein models need 125 to 227~GiB, while Algorithm~\ref{alg:acg} needs at most 58~GiB.}

\subsection{Discussion}
The choice of the set is a trade-off between reduction, computation time and accuracy on unseen loading. On the test day of South Alburg, only one of the 24 hours lies in the box and none in the convex hull. Both reductions exceed $\bar{E}$ in 8 hours, with errors up to $7.1\bar{E}$ with LinDistFlow and $7.3\bar{E}$ with AC power flow. The violations happen in 4 nodes, with the largest contribution from the load at the bus of the worst node, up to 17 times its largest training value. Projected onto their own sets, these hours satisfy the bound, as certified. With the more conservative ellipsoid and Wasserstein sets, the test-day errors stay below $\bar{E}$. The box gives a large reduction in minutes, but its accuracy depends on whether future injections stay in the sampled range.

\section{Conclusion}
This paper extended Opti-KRON to account for uncertain nodal injections. Under the LinDistFlow model, the proposed formulation enforces voltage magnitude error bounds over robust uncertainty sets and worst-case CVaR bounds under a Wasserstein ambiguity set. On the South Alburg feeder, the sets that stay within the sampled range of the injections gave the largest reductions. Constraint screening removed most error constraints before the solve. For one robust uncertainty set, only Algorithm~\ref{alg:acg} proved the minimal number of super-nodes, as the direct model ran out of memory.

The guarantee holds only for the approximate LinDistFlow errors and for injections in the set. On the test day, the largest reductions exceeded the voltage magnitude error threshold in hours where some loads lay far outside their training range. \rev{Future work will calibrate the size of the sets on separate data and for each voltage magnitude error, since each error depends mainly on a few injections. A margin on the threshold could cover the difference between the LinDistFlow and AC errors. We also plan to use the robust reductions in OPF problems, where the threshold tightens the voltage limits.}

\bibliographystyle{IEEEtran}
\bibliography{references}

\begin{thebibliography}{10}
\providecommand{\url}[1]{#1}
\csname url@samestyle\endcsname
\providecommand{\newblock}{\relax}
\providecommand{\bibinfo}[2]{#2}
\providecommand{\BIBentrySTDinterwordspacing}{\spaceskip=0pt\relax}
\providecommand{\BIBentryALTinterwordstretchfactor}{4}
\providecommand{\BIBentryALTinterwordspacing}{\spaceskip=\fontdimen2\font plus
\BIBentryALTinterwordstretchfactor\fontdimen3\font minus \fontdimen4\font\relax}
\providecommand{\BIBforeignlanguage}[2]{{%
\expandafter\ifx\csname l@#1\endcsname\relax
\typeout{** WARNING: IEEEtran.bst: No hyphenation pattern has been}%
\typeout{** loaded for the language `#1'. Using the pattern for}%
\typeout{** the default language instead.}%
\else
\language=\csname l@#1\endcsname
\fi
#2}}
\providecommand{\BIBdecl}{\relax}
\BIBdecl

\bibitem{Dorfler2013}
F.~D{\"o}rfler and F.~Bullo, ``Kron reduction of graphs with applications to electrical networks,'' \emph{IEEE Trans. Circuits Syst. I, Reg. Papers}, vol.~60, no.~1, pp. 150--163, 2013.

\bibitem{Pecenak2019}
Z.~K. Pecenak, V.~R. Disfani, M.~J. Reno, and J.~Kleissl, ``Inversion reduction method for real and complex distribution feeder models,'' \emph{IEEE Trans. Power Syst.}, vol.~34, no.~2, pp. 1161--1170, 2019.

\bibitem{Chevalier2022}
S.~Chevalier and M.~Almassalkhi, ``Towards optimal {Kron-based} reduction of networks {(Opti-KRON)} for the electric power grid,'' in \emph{2022 IEEE 61st Conference on Decision and Control (CDC)}, 2022, pp. 5713--5718.

\bibitem{Mokhtari2026}
O.~Mokhtari, S.~Chevalier, and M.~Almassalkhi, ``Structure-preserving optimal {Kron}-based reduction of radial distribution networks,'' \emph{IEEE Trans. Power Syst.}, 2026, early access, doi: \mbox{10.1109/TPWRS.2026.3726028}.

\bibitem{MokhtariPSCC2026}
O.~Mokhtari, S.~Chevalier, and M.~Almassalkhi, ``Optimal {Kron-based} reduction of networks ({Opti-KRON}) for three-phase distribution feeders,'' \emph{Electric Power Systems Research}, vol. 263, p. 113615, 2027.

\bibitem{Roald2023}
L.~A. Roald, D.~Pozo, A.~Papavasiliou, D.~K. Molzahn, J.~Kazempour, and A.~Conejo, ``Power systems optimization under uncertainty: {A} review of methods and applications,'' \emph{Electr. Power Syst. Res.}, vol. 214, 2023, {Art.} no. 108725.

\bibitem{Baran1989}
M.~E. Baran and F.~F. Wu, ``Network reconfiguration in distribution systems for loss reduction and load balancing,'' \emph{IEEE Trans. Power Del.}, vol.~4, no.~2, pp. 1401--1407, Apr. 1989.

\bibitem{Gan2014}
L.~Gan and S.~H. Low, ``Convex relaxations and linear approximation for optimal power flow in multiphase radial networks,'' in \emph{Proc. Power Syst. Comput. Conf. (PSCC)}, 2014, pp. 1--9.

\bibitem{BenTal2009}
A.~Ben-Tal, L.~El~Ghaoui, and A.~Nemirovski, \emph{Robust Optimization}.\hskip 1em plus 0.5em minus 0.4em\relax Princeton, NJ: Princeton University Press, 2009.

\bibitem{EsfahaniKuhn2018}
P.~Mohajerin~Esfahani and D.~Kuhn, ``Data-driven distributionally robust optimization using the {Wasserstein} metric: Performance guarantees and tractable reformulations,'' \emph{Math. Program.}, vol. 171, no. 1-2, pp. 115--166, 2018.

\bibitem{MutapcicBoyd2009}
A.~Mutapcic and S.~Boyd, ``Cutting-set methods for robust convex optimization with pessimizing oracles,'' \emph{Optim. Methods Softw.}, vol.~24, no.~3, pp. 381--406, 2009.

\bibitem{BertsimasDunningLubin2016}
D.~Bertsimas, I.~Dunning, and M.~Lubin, ``Reformulation versus cutting-planes for robust optimization,'' \emph{Comput. Manag. Sci.}, vol.~13, no.~2, pp. 195--217, 2016.

\bibitem{Schneider2018}
K.~P. Schneider \emph{et~al.}, ``Analytic considerations and design basis for the {IEEE} distribution test feeders,'' \emph{IEEE Trans. Power Syst.}, vol.~33, no.~3, pp. 3181--3188, May 2018.

\bibitem{Fan2013}
J.~Fan, Y.~Liao, and M.~Mincheva, ``Large covariance estimation by thresholding principal orthogonal complements,'' \emph{J. R. Stat. Soc. Ser. B Stat. Methodol.}, vol.~75, no.~4, pp. 603--680, 2013.

\bibitem{BertsimasSim2004}
D.~Bertsimas and M.~Sim, ``The price of robustness,'' \emph{Oper. Res.}, vol.~52, no.~1, pp. 35--53, 2004.

\end{thebibliography}

\end{document}